\documentclass[fleqn,usenatbib]{rasti}

\usepackage{newtxtext,newtxmath}

\usepackage[T1]{fontenc}

\DeclareRobustCommand{\VAN}[3]{#2}
\let\VANthebibliography\thebibliography
\def\thebibliography{\DeclareRobustCommand{\VAN}[3]{##3}\VANthebibliography}

\usepackage{siunitx}
\usepackage{float}
\usepackage{placeins}
\usepackage{minted}

\usepackage{caption}
\usepackage{booktabs}
\usepackage{multirow}

\usepackage{CJKutf8}
\usepackage{orcidlink}
\usepackage{ulem}

\usepackage{natbib}

\usepackage{graphicx}	
\usepackage{amsmath}	

\title[WEAVE-TwiLight-Survey]{The WEAVE-TwiLight-Survey: Expanding WEAVE’s Reach to Bright and Low-Surface-Density Targets with a Novel Observing Mode}

\author[T. Hajnik et al.]{Thomas Hajnik,$^{1}$\thanks{E-mail: th721@ast.cam.ac.uk}\orcidlink{0009-0001-0409-3019}
Nicholas A. Walton,$^{1}$\orcidlink{0000-0003-3983-8778}
Giuseppe D'Ago,$^{1}$\orcidlink{0000-0001-9697-7331}
Piercarlo Bonifacio,$^{2}$\orcidlink{0000-0002-1014-0635}
\newauthor{Gavin Dalton,$^{3,4}$\orcidlink{0000-0002-3031-2588}}
Lilian Domínguez-Palmero,$^{5,6}$\orcidlink{0009-0004-8470-9304}
Emanuel Gafton,$^{5}$\orcidlink{0000-0003-0781-6638}
Mike J. Irwin,$^{1}$\orcidlink{0000-0002-2191-9038}
Sergio Picó,$^{5}$
\newauthor{David Terrett,$^{3, 4}$}
Anke Ardern-Arentsen,$^{1}$\orcidlink{0000-0002-0544-2217}
Rubén Sánchez-Janssen,$^{5}$\orcidlink{0000-0003-4945-0056}
David S. Aguado,$^{6,7,1}$\orcidlink{0000-0001-5200-3973}
\newauthor{J. Alfonso L. Aguerri,$^{6,7}$\orcidlink{0000-0002-2839-2144}}
Carlos Allende Prieto,$^{6,7}$\orcidlink{0000-0002-0084-572X}
Marc Balcells,$^{5, 6, 7}$\orcidlink{0000-0002-3935-9235}
Chris Benn,$^{5}$
Angela Bragaglia,$^{8}$\orcidlink{0000-0002-0338-7883}
\newauthor{Elisabetta Caffau,$^{2}$\orcidlink{0000-0001-6011-6134}}
Esperanza Carrasco,$^{9}$\orcidlink{0000-0002-9174-5491}
Ricardo Carrera,$^{8}$\orcidlink{0000-0001-6143-8151}
Silvano Desidera,$^{10}$\orcidlink{0000-0001-8613-2589}
Boris T. G\"ansicke,$^{11}$\orcidlink{0000-0002-2761-3005}
\newauthor{Sarah Hughes,$^{12}$\orcidlink{0000-0002-7332-2751}}
Shoko Jin,$^{13}$\orcidlink{0000-0002-4824-8430}
Ian Lewis,$^{3}$\orcidlink{0009-0009-2090-5513}
Alireza Molaeinezhad,$^{1}$\orcidlink{0000-0002-2477-6634}
David N. A. Murphy,$^{1}$
\newauthor{Ellen Schallig,${^{3}}$\orcidlink{0000-0003-2517-0965}}
Scott Trager,$^{13}$\orcidlink{0000-0001-6994-3566}
and Antonella Vallenari $^{10}$\orcidlink{0000-0003-0014-519X}
\\
\\
$^{1}$Institute of Astronomy, University of Cambridge, Madingley Road, \mbox{Cambridge, CB3 0HA, U.K.}\\
$^{2}$LIRA, Observatoire de Paris, Universit{\'e} PSL, Sorbonne Universit{\'e}, Universit{\'e} Paris Cité, CY Cergy Paris Universit{\'e}, CNRS, 92190 Meudon, France\\
$^{3}$Oxford Astrophysics, University of Oxford, Keble Road, Oxford, OX1 3RH, U.K.\\
$^{4}$RALSpace, STFC Rutherford Appleton Laboratory, Harwell, OX11 0QX, U.K.\\
$^{5}$Isaac Newton Group, 38700 Santa Cruz de La Palma, Spain\\
$^{6}$Instituto de Astrof\'isica de Canarias, calle Vía L\'actea s/n, E-38205 La Laguna, Spain \\
$^{7}$Departamento de Astrof\'isica, Universidad de La Laguna, Avenida Astrof\'isico Francisco S\'anchez s/n, E-38206 La Laguna, Spain \\
$^{8}$INAF – Osservatorio di Astrofisica e Scienza dello Spazio di Bologna, via P. Gobetti 93/3, 40129, Bologna, Italy \\
$^{9}$Instituto Nacional de Astrofísica, Óptica y Electrónica, Luis Enrique Erro 1, C.P. 72840, Tonantzintla, Puebla, México \\
$^{10}$INAF – Osservatorio Astronomico di Padova, Vicolo dell’Osservatorio 5, 35122, Padova, Italy\\
$^{11}$Department of Physics, University of Warwick, Coventry, CV4 7AL, U.K. \\
$^{12}$MIT Kavli Institute for Astrophysics and Space Research, 77 Massachusetts Avenue, Cambridge, MA 02139, USA\\
$^{13}$Kapteyn Astronomical Institute, Rijksuniversiteit Groningen, Landleven 12, 9747 AD, Groningen, The Netherlands\\
}

\date{Accepted XXX. Received YYY; in original form ZZZ}

\pubyear{\the\year{}}

\begin{document}
\label{firstpage}
\pagerange{\pageref{firstpage}--\pageref{lastpage}}
\maketitle

\begin{abstract}
Current-day multi-object spectroscopic surveys are often limited in their ability to observe bright stars due to their low surface densities, resulting in increased observational overheads and reduced efficiency. Addressing this, we have developed a novel observing mode for WEAVE (William Herschel Telescope Enhanced Area Velocity Explorer) that enables efficient observations of low-surface-density target fields without incurring additional overheads from calibration exposures. As a pilot for the new mode, we introduce the WEAVE-TwiLight-Survey (WTLS), focusing on bright exoplanet-host stars and their immediate surroundings on the sky. High observational efficiency is achieved by superimposing multiple low-target-density fields and allocating the optical fibres in this configuration. We use a heuristic method to define fields relative to a central guide star, which serves as a reference for their superposition. Suitable guide fibres for each merged configuration are selected using a custom algorithm. Test observations have been carried out, demonstrating the feasibility of the new observing mode. We show that merged field configurations can be observed with WEAVE using the proposed method. The approach minimizes calibration times and opens twilight hours to WEAVE's operational schedule. WTLS is built upon the new observing mode and sourced from the ESA \textit{PLATO} long-duration-phase fields. This survey will result in a homogeneous catalogue of $\sim6\,300$ bright stars, including 62 known planet hosts, laying the groundwork for future elemental abundance studies tracing chemical patterns of planetary formation. This new observing mode (WEAVE-Tumble-Less) expands WEAVE's capabilities to rarely used on-sky time and low-density field configurations without sacrificing efficiency.
\end{abstract}




\section{Introduction}\label{se:intro}
We live in the age of exoplanet discoveries, and over the past few years the field has seen rapid advancements in detecting (e.g. \mbox{\citealp{boruckiKeplerPlanetDetectionMission2010}}; \citealp{rickerTransitingExoplanetSurvey2014}; \citealp{rauerPLATOMission2025}) and characterizing planets (e.g. \citealp{adibekyanCompositionalLinkRocky2021}; \citealp{bonsorHoststarExoplanetCompositions2021}, \mbox{\citealp{madhusudhanCarbonbearingMoleculesPossible2023}}) as well as their host stars. Modern exoplanet detection missions, such as TESS \citep{rickerTransitingExoplanetSurvey2014} and \textit{PLATO} \citep{rauerPLATOMission2025}, provide the most accurate stellar and planetary parameters for stars with $V \leq 11$ mag. To better understand the exoplanet population in terms of occurrence rates, planet types, system architectures, and the potential effects of the immediate stellar and Galactic neighbourhood, large and homogeneous spectral datasets are essential for enabling meaningful statistical investigation. \par
\begin{figure}
\centering
 \resizebox{\hsize}{!}{\includegraphics{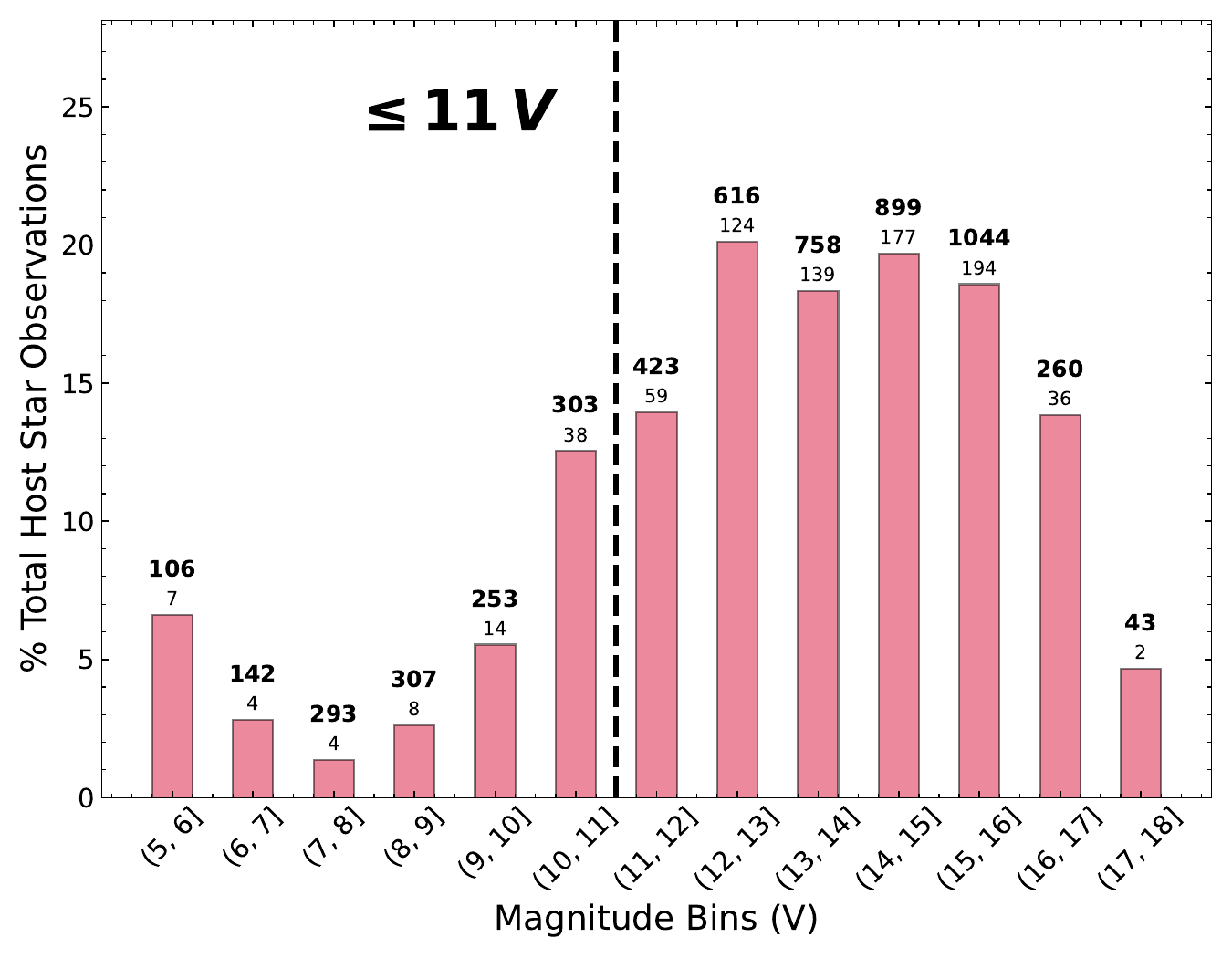}}
   \caption{Fraction of currently known host stars (NASA Exoplanet Archive, August 2024) that have been observed by APOGEE DR17 \mbox{\citep{abdurroufSeventeenthDataRelease2022}}, grouped by magnitude bin. The total number of known exoplanet hosts is shown in bold above each bin. The number below indicates the number of observed hosts in APOGEE DR17 in that bin. Note the sharp decline in observations for magnitudes brighter than 11 (to the left of the dashed line). This reflects the low sky densities of bright stars, which result in higher observational overheads.}
  \label{im:obs_per_mag}
\end{figure}

However, the observation of bright stars ($V \leq 11$ mag) comes with significant observational overheads for multi-object spectroscopy surveys, due to their low on-sky number densities, compared to stars at fainter magnitudes. This is apparent when plotting the fraction of currently known exoplanet hosts that are part of APOGEE DR17 (APOGEE hereafter, \citealp{abdurroufSeventeenthDataRelease2022}), grouped by their respective magnitudes (see Fig.~\ref{im:obs_per_mag}). The sharp decline in the number of observations for targets $V \leq 11$ mag, apparent in Figure \ref{im:obs_per_mag}, is due to said operational overheads. In addition, the large differences in required exposure times between bright and faint targets renders simultaneous observations inefficient, as avoiding saturation of the bright sources within the field of view demands many short exposures. This increases the total detector readout time and reduces overall efficiency. \par
The above underscores the need for efficient, large-scale multi-object spectroscopic follow-up surveys to build up a representative sample of bright host stars. An important additional benefit of such surveys is the ability to identify and reject false positives, such as spectroscopic binaries, which can produce radial velocity signatures that mimic those of planetary companions. \par

To address this, a low-density/bright-star mode has been developed for the new WHT (William Herschel Telescope) Enhanced Area Velocity Explorer (WEAVE: \citealp{daltonWEAVENextGeneration2012}, \citeyear{daltonFinalDesignProgress2016}; \mbox{\citealp{jinWidefieldMultiplexedSpectroscopic2024}}). This novel observing technique combines multiple observational fields into a single fibre configuration, enabling rapid sequential observations without reconfiguring the fibres. The approach avoids additional calibration exposures between pointings and has been successfully tested for field offsets of up to $\sim$$15^\circ$, significantly enhancing observing efficiency at low surface densities. \par
The new setup acts as the foundation for the WEAVE-TwiLight-Survey (WTLS), focusing on bright exoplanet host stars and their immediate surroundings on the sky ($\sim 2$-$5^\circ$ deg). We introduce this pilot survey to enable evaluation of the new mode during daily operations while generating a homogeneous, high-precision dataset of chemical abundances and stellar parameters, for approximately $6\,300$ stars with magnitudes in the range $6 \leq V \leq 11.5$. WTLS is primarily defined as a twilight survey in order to minimize disruption to ongoing and planned WEAVE surveys, as well as open-time programmes. However, we believe that the novel approach has great potential for night-time implementation, for reasons provided later in this paper.\newline

In Section \ref{se:twilight_obs_mode} of this work, we focus on the need for a low-target-density observing mode for WEAVE, followed by a detailed description of the WEAVE-TwiLight-Survey input catalogue (see Section \ref{s:defining_fields_for_WTL}), the creation of merged field configurations (Section \ref{ss:field_definition_guide_alloc}) and its final on-sky fields in Section \ref{s:WTL_fields}. A preliminary look into data from recent test observations is given in Section \ref{s:test_observations}. We conclude with a short summary and future outlook in Section \ref{s:conclusion}.

\section{A low-density and bright star observing mode for WEAVE}\label{se:twilight_obs_mode}
WEAVE is optimized to observe targets in the magnitude range \mbox{$12 \leq V \leq 20$}  with numerous sources per field, resulting in approximately $1\,000$ spectra for each observing block (OB) in its multi-object-spectroscopy (MOS) mode. MOS offers a low- and high-resolution observing mode with a spectral resolution of $\sim5\,000$ and $\sim20\,000$, respectively. High-resolution data can be gathered using either the \textit{blue1} or \textit{blue2} spectrograph mode (see \mbox{\citealp{jinWidefieldMultiplexedSpectroscopic2024}}), with \textit{blue1} covering $4040$\,\AA{} to $4650$\,\AA{} and \textit{blue2} covering $4730$\,\AA{} to $5450$\,\AA{}. Both settings come with additional coverage towards longer wavelengths (\textit{red}) from $5950$\,\AA{} to $6850$\,\AA{}. In this paper, we exclusively focus on the fibre-fed multi-object-spectroscopy high-resolution mode (MOS HR).
For additional information on other setups (e.g. low resolution and IFU) and WEAVE's survey strategy, we refer to \cite{jinWidefieldMultiplexedSpectroscopic2024}. A technical overview of WEAVE can be found in \cite{daltonFinalDesignProgress2016}. Data recorded with WEAVE is subsequently processed using the Core Processing System (CPS), responsible for wavelength calibration, telluric correction, sky subtraction and stacking of individual exposures. Next, the data are fed to WEAVE's Advanced Processing System (APS), handling target classification and parameter inference. For stellar data this entails the inference of radial velocities, main stellar parameters ($T_{\mathrm{eff}}$, $\log g$, [Fe/H], [$\alpha$/H] and $v_{\text{micro}}$) and in some cases atmospheric chemical abundances. \newline

As mentioned above, WEAVE is optimized to observe targets with $V\geq 12$ mag. However, for sources brighter than $V=11$ mag, the average on-sky number density drops to roughly $20$ stars deg$^{-2}$ \mbox{(e.g. \citealp{zakharovMinimumStarTracker2013})}. In contrast, stars of magnitude $V=12.5$ reach densities of about $100$ stars deg$^{-2}$. This sharp decline significantly reduces observational efficiency. Before presenting a new observing mode addressing this problem, a brief technical overview of regular WEAVE observations and the critical components involved is given below. 

\subsection{Overview of Regular WEAVE Observations}
WEAVE has a circular field of view, covering an on-sky area of  $\sim\pi$ deg$^2$ at the telescope's prime focus. Inside the fibre positioner unit, a field plate is mounted in the image plane of the focus corrector. Two Cartesian robots place optical fibres onto this plate, aligning them precisely with the light paths of individual targets. The instrument employs two separate field plates (plates A and B) that are used interchangeably, one observing while the other undergoes fibre configuration. The plates are mounted on opposite ends of a cylindrical structure that can be rotated, allowing the fields to alternate between observing and fibre-placement positions, a process referred to as `tumbling'. \par
WEAVE utilizes two types of optical fibres: \textit{science} and \textit{guide}. The first are standard MOS (multi-object spectroscopy) fibres, subtending 1.3\si{\arcsecond} on the sky, and are used for both science and calibration measurements \citep{jinWidefieldMultiplexedSpectroscopic2024}. Guide fibres are coherent imaging bundles with a diameter of 4.5\si{\arcsecond}, feeding signals to the CCD of the autoguider system and assisting in target acquisition and pointing correction during exposure. At least three guide fibres per field are required to ensure successful observations.

To find an optimal fibre configuration, WEAVE employs the software \textsc{configure} \mbox{\citep{terrettFibrePositioningAlgorithms2014}}. Within its design, \textsc{configure} supports multiple fields per configuration to accommodate dithering strategies for WEAVE's Integral Field Unit modes (LIFU and mIFU, see \citealp{jinWidefieldMultiplexedSpectroscopic2024}). This feature is key to our newly developed observing strategy, outlined in Section~\ref{s:defining_fields_for_WTL}. WEAVE's standard observing mode leads to one observed field per pointing. Once the exposures are complete, the field plate cylinder is rotated to bring the alternate field plate into position. The tumbling procedure takes about $90$\,\si{\second} and results in a fibre slit change, requiring new calibration exposures to correct for possible variations in slit positioning repeatability. In this scenario, high observing efficiency is achieved by maximizing the number of targets per field, allowing a large number of measurements to be taken at each pointing, keeping observational overheads at a minimum. 

\subsection{Addressing Low-Surface-Density Field Constraints}
The above has a significant negative impact on the efficiency with which low-surface-density fields (e.g.\ bright stars) can be observed, as each new field requires tumbling the positioner and additional calibration exposures, yet yields relatively few targets ($\sim30$) compared to high-density OBs ($\sim1\,000$). This inevitably leads to a preference for fainter sources. 
To avoid saturation at the bright end of the observed magnitude range, pointings should be split into multiple exposures. Although it is theoretically possible to observe targets spanning several orders of magnitude on the same plate, the increasing number of exposures required to prevent saturated pixels introduces additional overheads due to readout times. For WTLS, the adopted magnitude range ($6 \leq V \leq 11.5$) was successfully tested by exposing each test field three times (see Section \ref{s:test_observations}). Except for a very small subset of WTLS fields, the brightness range within a single pointing spans $2.5$ magnitudes on average. In merged field configurations of bright targets, cross-talk remains minimal because their low surface densities make it unlikely for two stars to occupy adjacent slit positions. In such cases, the measured cross-talk is $\sim 1$ percent at the detector centre. For higher-density fields, exposure times and magnitude ranges must be selected more carefully. \par

To overcome the inefficiencies described above and extend WEAVE’s capabilities to brighter magnitudes, a new operational mode, WEAVE-Tumble-Less (WEAVE-TL), has been developed. It employs a novel OB-preparation and observing strategy that allows multiple fields to be set up and observed with a single fibre configuration. This is enabled by superimposing (i.e. merging) individual fields in reference to a central guide star (see Section \ref{ss:field_definition_guide_alloc}), creating a merged field configuration, referred to as a `field-batch' hereafter. The two limiting factors of this technique are the availability of operational guide fibres and the distribution of guide stars in the fields, both of which must be taken into account in the field definition process. \\

For the WEAVE-TL mode, both \textsc{configure} \citep{terrettFibrePositioningAlgorithms2014} and the Observatory Control System (OCS, \mbox{\citealp{picoWEAVEObservatoryControl2018}}) software have been adapted to allow the telescope to move to a new set of targets without requiring the fibre positioner to tumble. Only data from the guide fibres relevant to each field are sent to the autoguider software at each pointing, enabling field acquisition and guiding. Test observations have shown that WEAVE-TL OBs are on average 20 percent faster to execute than observing the same fields using the regular mode.

\section{The WEAVE-Twilight-Survey Input Catalogue}\label{s:defining_fields_for_WTL}
\begin{figure*}
\centering
 \includegraphics[width=17cm]{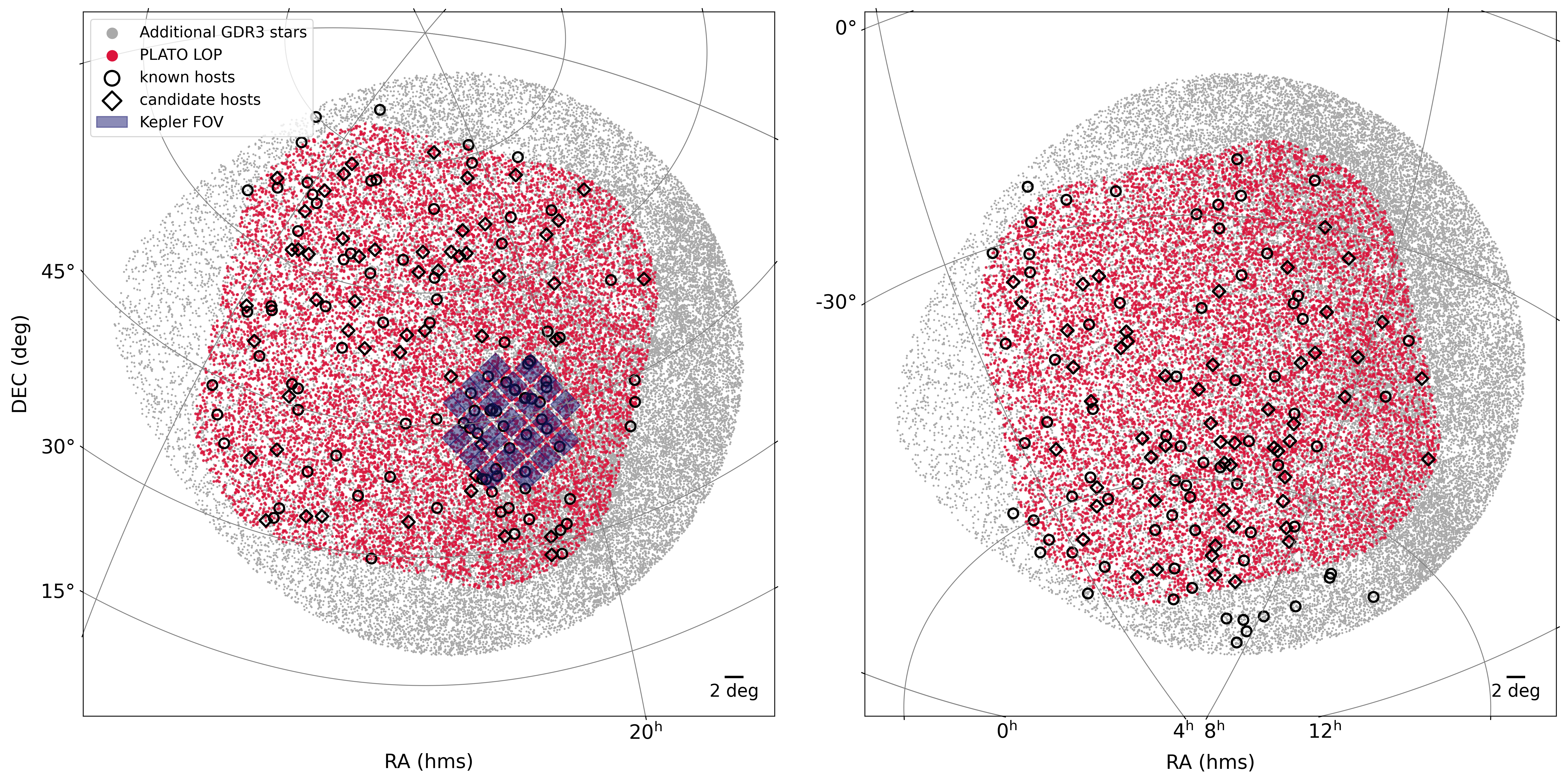}
   \caption{Sky projection of the \textit{proto} catalogues after applied magnitude cuts (see Section \ref{s:defining_fields_for_WTL}) in Equatorial coordinates, constituting the superset from which the northern and southern WEAVE-TwiLight fields are derived. \textit{Red} stars are part of the \textit{PLATO} Long-duration Observation Phase (LOP) fields \mbox{(\citealp{nascimbeniPLATOFieldSelection2022}, \citeyear{nascimbeniPLATOFieldSelection2025})}. \textit{Black} circles and diamonds signify known exoplanet hosts and candidates, respectively. For orientation the Kepler FOV is overlaid in \textit{dark blue}.}
  \label{im:initial_fields}
\end{figure*}
The starting points for the WEAVE-TwiLight input catalogue are the northern and southern \textit{PLATO} Long-duration Observation Phase (LOP) fields (LOPN and LOPS, see Table \ref{t:LOP_fields}). These fields are derived from \textit{Gaia} \citep{gaiacollaborationGaiaMission2016} stars and subject to filtering according to \cite{montaltoAllskyPLATOInput2021} and \mbox{\cite{nascimbeniPLATOFieldSelection2022, nascimbeniPLATOFieldSelection2025}}. \citeauthor{montaltoAllskyPLATOInput2021} select FGK and M stars ($V \leq 16$ mag) using cuts in reddening-corrected colour-magnitude diagrams, informed by Galactic simulations and observations. To ensure detectability, the authors refine their sample to meet \textit{PLATO's} photometric noise requirements (e.g. $< 50$\,ppm/hr) and derive stellar parameters ($T_{\mathrm{eff}}$, radius and mass), in order to minimize contamination and metallicity bias. As WTLS is planned as a twilight survey, the \textit{PLATO} LOP fields were restricted in visual magnitude to stars fulfilling \mbox{$6 \leq V \leq 11.5$ mag} (corresponding to approximately $5.8 \leq G_{\mathit{Gaia}} \leq 10.6$ mag). For this, F, G and K stars, as defined by the \textit{PLATO} input catalogue parameter \texttt{sourceFlag} (version \texttt{PIC2.0}) are selected. To further enhance the final target yield, we draw a circle with a radius of 36\si{\degree} in Galactic coordinates centred on each LOP field. This circle is then filled with stars from the \textit{Gaia} archive, covering magnitudes  from $6 \leq G_{\mathit{Gaia}} \leq 10$, with the following flags: \texttt{parallax\_over\_error} $<5$, absolute \texttt{pmra} and \texttt{pmdec} $<50$ mas yr$^{-1}$. This ensures good quality parallax measurements and excludes high-proper-motion stars. The result are two catalogues (\textit{proto} catalogues hereafter), from which the WEAVE-TwiLight fields are subsequently derived (see Fig.~\ref{im:initial_fields}). Here, the footprint of \textit{PLATO}'s LOPN and LOPS fields can be seen in red. \par
\begin{table}
    \centering
    \resizebox{\columnwidth}{!}{%
    \begin{tabular}{ccccc} \toprule
         LOP field &  $\alpha$ (ICRS)&  $\delta$ (ICRS)& $l$ (IAU 1958)& $b$ (IAU 1958)\\ \midrule
         north &  277.18023\si{\degree}& 52.85952\si{\degree} & 81.56250\si{\degree}  & 24.62432\si{\degree} \\
         south &  95.31043\si{\degree}&  $-$47.88693\si{\degree}&  255.93750\si{\degree}& $-$24.62432\si{\degree} \\ \bottomrule \\
    \end{tabular}
    }
    \caption{Field centre coordinates for \textit{PLATO}'s Long-duration Observation Phase (LOP) fields taken from (\protect\citealp{nascimbeniPLATOFieldSelection2022}, \protect \citeyear{nascimbeniPLATOFieldSelection2025}). The WEAVE-TwiLight fields share these centre coordinates and extend the \textit{PLATO} fields using bright stars from \textit{Gaia} DR3. Please note that only a small fraction of the southern LOP targets are accessible to the WHT.}
    \label{t:LOP_fields}
\end{table}

In Figure \ref{im:initial_fields}, known planet hosts and and candidates are depicted with black circles and diamonds, respectively, by crossmatching the LOP fields with the TESS Objects of Interest (TOI) catalogue \mbox{\citep{nexsciExoplanetFollowupObserving2022}} and the \textit{STELLAR HOSTS} table\footnote{\url{https://exoplanetarchive.ipac.caltech.edu/cgi-bin/TblView/nph-tblView?app=ExoTbls&config=STELLARHOSTS}} from the NASA Exoplanet Archive \citep{akesonNASAExoplanetArchive2013a}. Sources in the crossmatch that had a TOI \texttt{TFOPWG Disposition} flag equal to \texttt{"PC"} were deemed host candidates. Sources with a flag value \texttt{"KP"} were marked as known planetary hosts. To maximize their number, a second crossmatch with the NASA Exoplanet Archive \textit{PLANETARY SYSTEMS} dataset\footnote{\url{https://exoplanetarchive.ipac.caltech.edu/cgi-bin/TblView/nph-tblView?app=ExoTbls&config=PS}} was performed. After dropping all hosts stars with a controversial planet flag (\texttt{pl\_controv\_flag}) value equal to 1, the northern and southern \textit{proto} catalogues contain 92 and 62 confirmed exoplanet hosts, respectively. Note that due to the magnitude cuts mentioned above, these numbers are smaller than in the original \textit{PLATO} LOP fields. \newline

To enable the new observing mode, individual WTLS fields (pointings) are defined with respect to suitable central and off-centre guide stars, as detailed in the following sections (\ref{sss:field_definition} and \ref{sss:guide_fibre_alloc}).

\section{Creating merged field configurations}\label{ss:field_definition_guide_alloc}
The ability to create merged field configurations is entirely dependent on the availability of usable guide stars in the selected input set. Especially during twilight, the possible field coverage is a function of the on-sky distribution of guide stars that are bright enough to be distinguished from the background sky signal. As a result, the field definition process is inherently connected to this distribution. This is further complicated by sky brightness variations of several magnitudes during twilight, as detailed below. To mitigate this, while also ensuring optimal field distribution and science target yield, a subset of bright stars from the WEAVE guide catalogue with magnitudes $14 \leq V \leq 15$ mag is added to the input set. \par

This range satisfies the brightness constraints of the guiding camera and ensures that individual guide stars differ by no more than one magnitude, enabling accurate acquisition and guiding. Measurements taken at the WHT between April 2024 and April 2025 indicate that the mean $V$-band surface brightness during twilight (civil, nautical, and astronomical) ranges from $\mu_V \sim 16$ to $22$~mag arcsec$^{-2}$. Given the relatively bright guide stars ($V \sim 14.5$ mag) used for WTLS, and all science targets falling in between $V=6$ and $11.5$ magnitudes, observations can begin shortly after sunset with good sky subtraction still being possible. \par

To enable the new observing mode, each of the individual fields need to be centred on one of the guide stars to enable their superposition. As mentioned above, every field in a batch must contain three guide stars in total. This requires two off-centre (secondary) guide stars per field in addition to the central one. Unlike the central guide fibre, the fibres placed on secondary guide stars cannot be shared across a field-batch. To accommodate these constraints, the creation of field-batches is divided into two steps: field definition and guide fibre allocation.

\subsection{Field Definition}\label{sss:field_definition}
The first problem is mathematically akin to a geometric set cover problem, where the largest possible set of points must be covered using a \textit{minimum number of circles}. In this particular instance, the circle centres must be placed solely on available guide stars. Mathematically, this can be written as
\begin{equation}
    S \supseteq G, T \quad \text{with} \quad G\neq T,
\end{equation}
where $S$ is the total set of stars, with $G$ being guide stars added to the science target catalogue $T$. Further, we let 
\begin{equation}
\begin{split}
        S &= \{s_{i} \mid i=1,2,...n\} \\
        G &= \{g_{j} \mid j=1,2,...m\}, \quad n,m \in \mathbb{N},
\end{split}
\end{equation}
where $s_{i}=(RA_{i}, DEC_{i})$ and $g_{j}=(RA_{j}, DEC_{j})$. The set of circles with radius $r$,  around every guide star in $S$ is defined as
\begin{equation}
        C = \{C_{k} \mid k = 1,2,...l\}, \quad l \in \mathbb{N},
\end{equation}
with each circle given as
\begin{equation}\label{eq:circle_constraint_1}
        C_{k}(g_{j}, r) = \{s_{i} \mid d(g_{j}, s_{i}) < r\},
\end{equation} 
where $d(g_{j}, s_{i})$ is the on-sky angular distance between a central guide star $g_{j}$ and a target star $s_{i}$. In the ideal case, the solution lies in finding the smallest subset $C^\prime \subseteq C$ that covers the whole of $S$, therefore
\begin{equation}
\begin{split}
        \text{minimize }\quad &|C^\prime|\\
        \text{subject to}\quad &\bigcup_{C_{k} \in C^\prime} C_{k} = S.
\end{split}
\end{equation}
The real world condition is less strict, as due to the on-sky distribution of stars and brightness constraints, perfect coverage is difficult to achieve. It can thus be written as
\begin{equation}
    \bigcup_{C_{k} \in C^\prime} C_{k} \lesssim S.
\end{equation}
Since it is not possible to completely cover a surface with non-overlapping circles, the offset parameter $\alpha \in \mathbb{R^+}$ is introduced. It can be controlled by the user and regulates the allowed distance between individual circle centres, maximising the overall number of stars in the fields. Letting $g_{i}, g_{j}$ $(i \neq j)$ be the centres of $C_{k}$ and $C_{l}$ $(k \neq l)$, the overlap can be constrained by
\begin{equation}\label{eq:overlap_condition}
     d(g_{i}, g_{j}) \geq \alpha.
\end{equation}
Incorporating the above into equation \ref{eq:circle_constraint_1} we find
\begin{equation}
C_{k}(g_{j}, r) = \{s_{i} \mid d(g_{j}, s_{i}) < r ~ \land ~  d(g_{i}, g_{j}) \geq \alpha   \}.
\end{equation}
Changing $\alpha$ allows fine-tuning the outcome by restricting the number of circles that will be placed onto the set.

To solve the above for the best possible configuration while keeping computational times minimal, a heuristic approach has been chosen, using a so-called \texttt{greedy} algorithm (e.g. \citealp{curtisClassificationGreedyAlgorithms2003}). First, the guide stars are sorted in descending order of brightness, ensuring that only the brightest ones are used as centre points. The algorithm then places a 1\si{\degree} radius circle around the brightest guide star and checks for overlaps with other circles. In the first iteration, this check will always pass, and the circle, including all stars within, becomes the first field. In subsequent iterations, new circles are placed around the remaining guide stars, again checking for the overlap condition given by Equation~\ref{eq:overlap_condition}. If a new circle violates this constraint, the algorithm moves to the next iteration. Each time a circle satisfies the condition, it becomes a new field, and its stars are removed from the remaining set. The algorithm terminates when no stars remain in the initial set. Using this method, an overall catalogue coverage of $\sim97$ percent has been achieved. \newline

Subsequently, individual fields are grouped into field-batches based on their proximity, resulting in up to three fields per batch. This number is limited by the availability of guide fibres. At the time the WTLS fields were defined, seven of the eight guide fibres on each field plate were operational. The number of operational guide fibres has since increased, as all previously broken fibres have been repaired. This will improve the coverage of merged field configurations in future iterations of WTLS, as discussed in Section \ref{sss:dependence_fibres_densities}. By defining fields around a central guide star, it is now possible to overlay all fields in a batch using their centre as a reference point.
\begin{figure}
\centering
 \resizebox{\hsize}{!}{\includegraphics{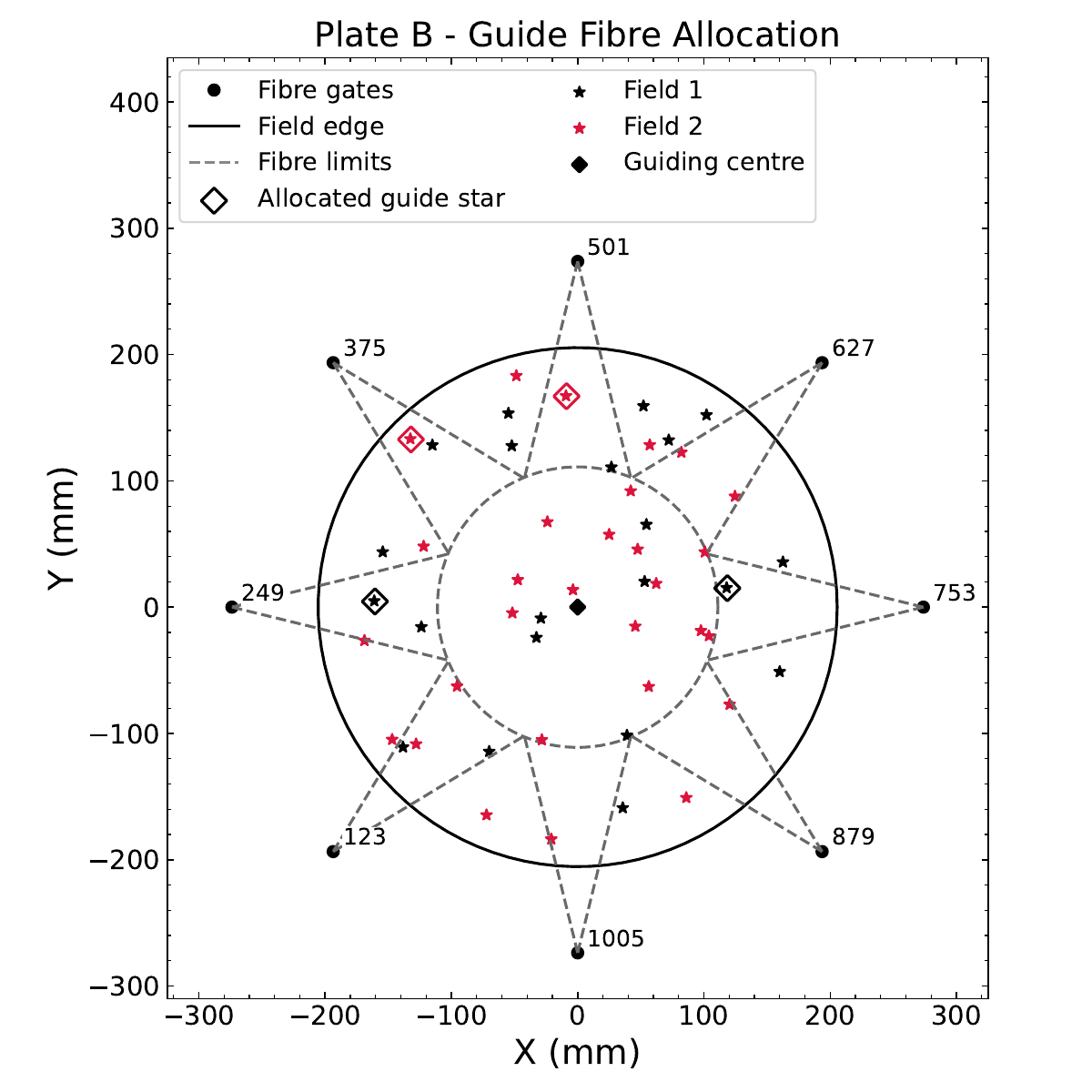}}
   \caption{Regions accessible by guide fibres are indicated by \textit{dashed} lines. Note that the allocation process used for WEAVE-TwiLight restricts the fibre reach to the end of each \textit{dashed} cone to prevent fibre crossings. The numbers along the fibre gates (\textit{black} dots) represent the fibre IDs for individual guide fibres on plate B. For plate A, the geometry remains the same. At the time the WTLS fields were defined, the following guide fibres were not in operation, plate A: 879; \mbox{plate B: 249.}}
  \label{im:guide_fibre_alloc}
\end{figure}
\begin{figure*}
\centering
 \includegraphics[width=17cm]{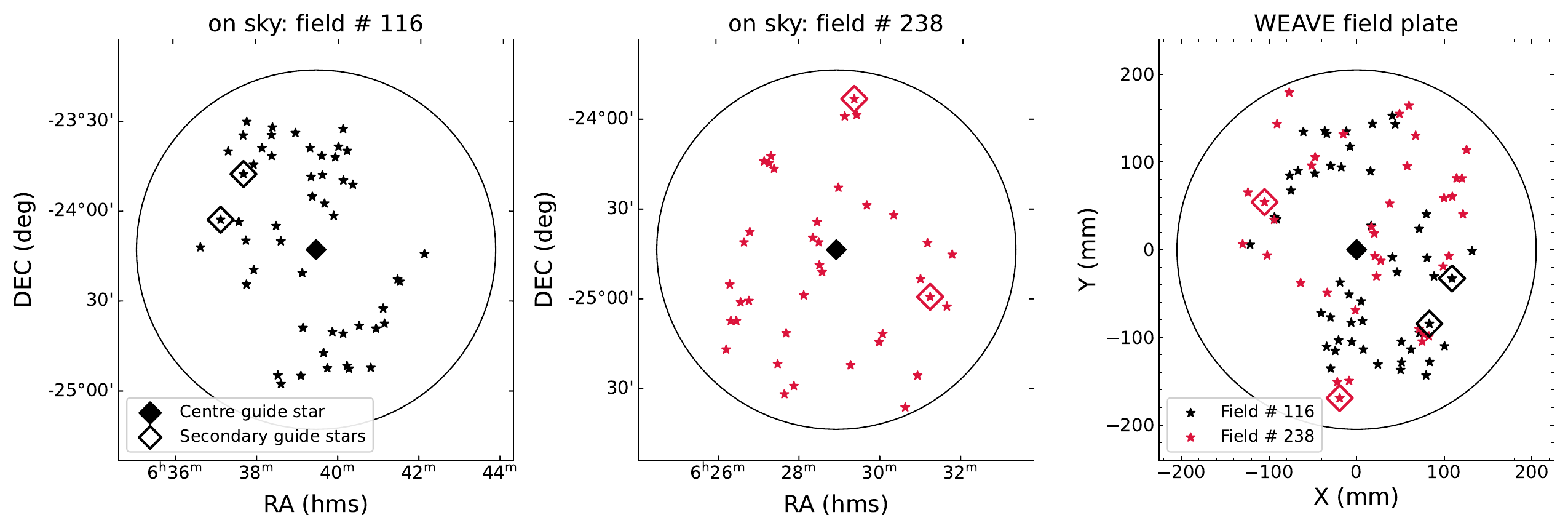}
   \caption{Merging of two fields on the sky onto the WEAVE field plate. Filled and empty diamonds (\textit{black \& red}) depict central and off-centre (secondary) guide stars, respectively. \textit{Left} and \textit{centre} panels: on-sky fields from the WEAVE-TwiLight-Survey input catalogue. \textit{Right:} the mirrored projection onto the field plate.}
  \label{im:field_merging}
\end{figure*}
\begin{figure*}
\centering
 \includegraphics[width=17cm]{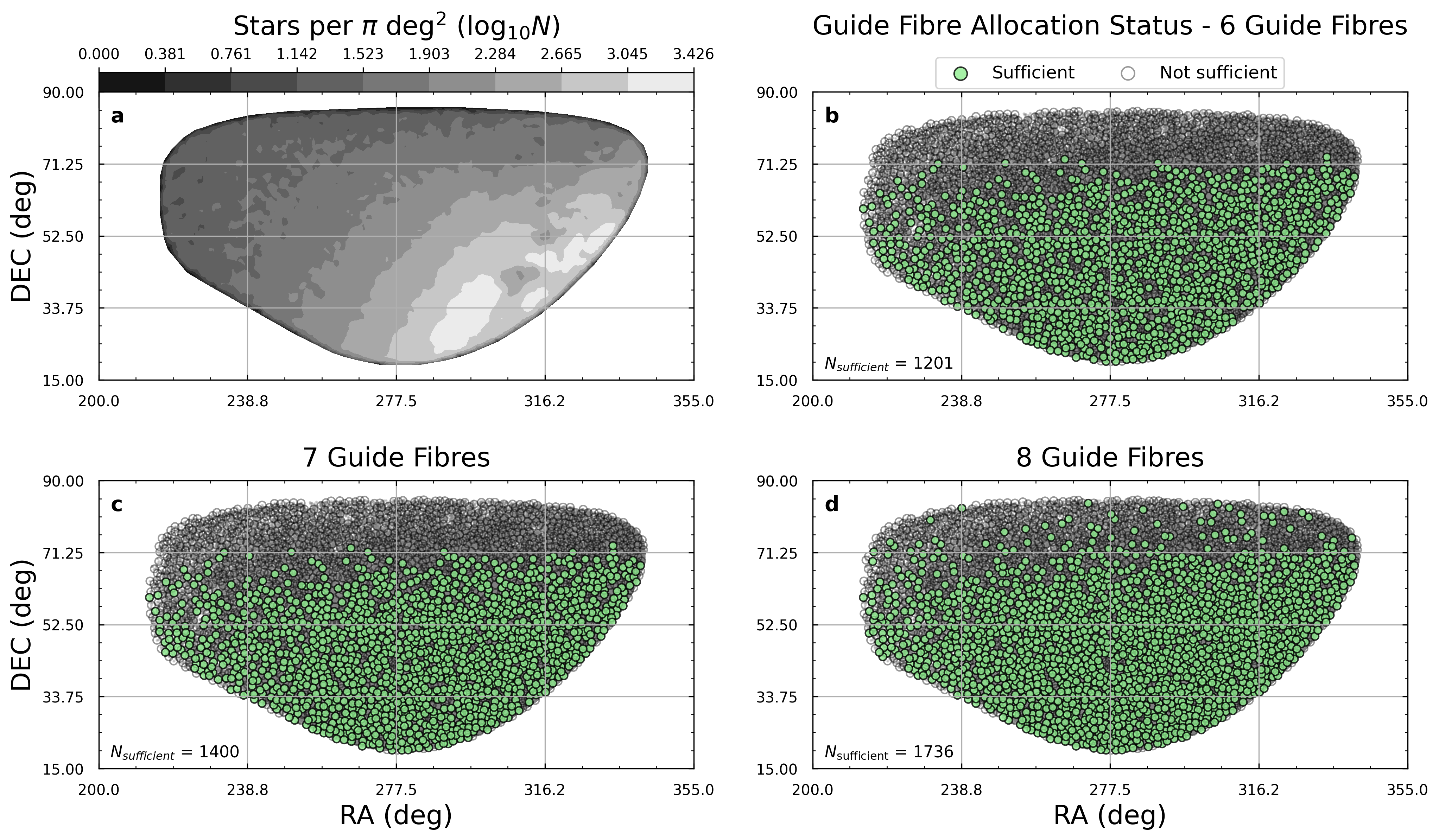}
   \caption{\textit{Panel a:} guide star number density per $\pi$ deg$^2$ ($G_{\mathit{Gaia}} \sim 14.5$ mag) in the northern WEAVE-TwiLight \textit{proto} field (see Fig.~\ref{im:initial_fields}, \textit{left} panel). \textit{Panels b, c and d:} \textit{green} circles show WEAVE pointings for which guide fibres could be successfully assigned (one central and two off-centre guide stars), allowing for autoguiding. \textit{Empty} circles depict fields where the algorithm was not able to determine a viable allocation. In the \textit{bottom left} corner of each panel, the number of fields with sufficient guide coverage ($N_{\mathrm{sufficient}}$) is given. Note that the above represents allocation results under twilight conditions. At nighttime, the number of viable guide stars increases significantly, allowing for field coverage of up to $\sim95$\% with 8 guide fibres.}
  \label{im:density_guide_allocation}
\end{figure*}

\subsection{Guide Fibre Allocation}\label{sss:guide_fibre_alloc}
In each field-batch, off-centre guide fibres must not be shared between fields, and therefore guide fibre IDs have to be assigned before the science fibres can be configured. WEAVE has two field plates onto which fibres can be deployed, each with 8 designated guide fibres. The geometry of the problem for field plate B is shown in \mbox{Figure \ref{im:guide_fibre_alloc}}. Note that the geometry remains the same for plate A. The guide fibre gates (black dots) are located around the field plate at angles 0 to 2$\pi$ in $\pi/4$ rad intervals. Each fibre can reach a region $\pm13.8$\si{\degree} from its gate up to the field centre (grey, dashed lines). This is further restricted to the region outside the grey dashed circle, in order to provide a better acquisition and guiding solution while avoiding the crossing of off-centre guide fibres.

The numbers next to the fibre gates correspond to their respective fibre IDs. A custom algorithm has been developed to solve for a valid configuration of guide fibres. The algorithm first assigns stars within reach of the guide fibres the fibre IDs of those guide fibres. If a star is only assigned one fibre ID, it retains this ID, which is then removed from the pool of available IDs. If multiple stars are assigned the same fibre ID, then the brightest star is selected for that guide fibre. After each successful allocation, the list is updated to reflect the current number of available guide fibres. The process continues until the number of fibres in the list does not change with respect to the previous allocation. This corresponds to either a completely allocated field-batch (one central and two off-centre guide fibres for each field) or to a situation where no remaining guide fibre can reach a guide star, resulting in an insufficiently allocated field. In the final step, each field-batch is assigned a central guide fibre from the remaining set of guide fibres, which is then shared between the individual fields. \par
An example of a merged field configuration with allocated guide fibres can be seen in Figure \ref{im:field_merging}. Here, diamonds depict the allocated central (black, filled-in) and secondary (black/red) guide stars. The rightmost panel shows the superimposed fields on the WEAVE field plate, taking into account projection and mirroring effects. 
Fields with sufficient guide fibre allocations are then handed to \textsc{configure}, which assigns science fibres to each target, completing the field configuration.

\subsection{Dependence on Target Number Density and Available Guide Fibres}\label{sss:dependence_fibres_densities}
The result of the method outlined above is a function of the location of the fields with respect to the Galactic plane and the number of operational guide fibres for a given field plate. For fields located high above the plane, the number of available guide stars is lower, as the overall number density of stars per $\pi$ deg$^2$ decreases in this direction. Therefore, fewer suitable guide stars become available. Additionally, the number of operational guide fibres significantly impacts the number of fields that can be successfully observed using the new mode. \par
Both above-mentioned aspects become apparent when examining Figure \ref{im:density_guide_allocation}, where the guide star number density per $\pi$\,deg$^2$ (panel a, upper-left) is compared with the guide fibre allocation status for each field under varying numbers of available guide fibres (remaining panels). Fields with sufficient guide fibre allocation (one central, two off-centre) are shown as green circles, while empty circles represent fields where the procedure was unsuccessful. It is clearly evident that the availability of more guide fibres significantly improves field coverage. This can be further improved by relaxing the brightness constraints on the guide star set, although this is not practical in twilight conditions, where stars brighter than the sky background are essential for target acquisition and guiding. When considering fainter magnitudes, the availability of more guide stars per unit area is expected to significantly improve guide fibre allocation. This is especially important for potential night-time surveys employing the WEAVE-Tumble-Less mode.
\begin{figure*}
\centering
 \includegraphics[width=17cm]{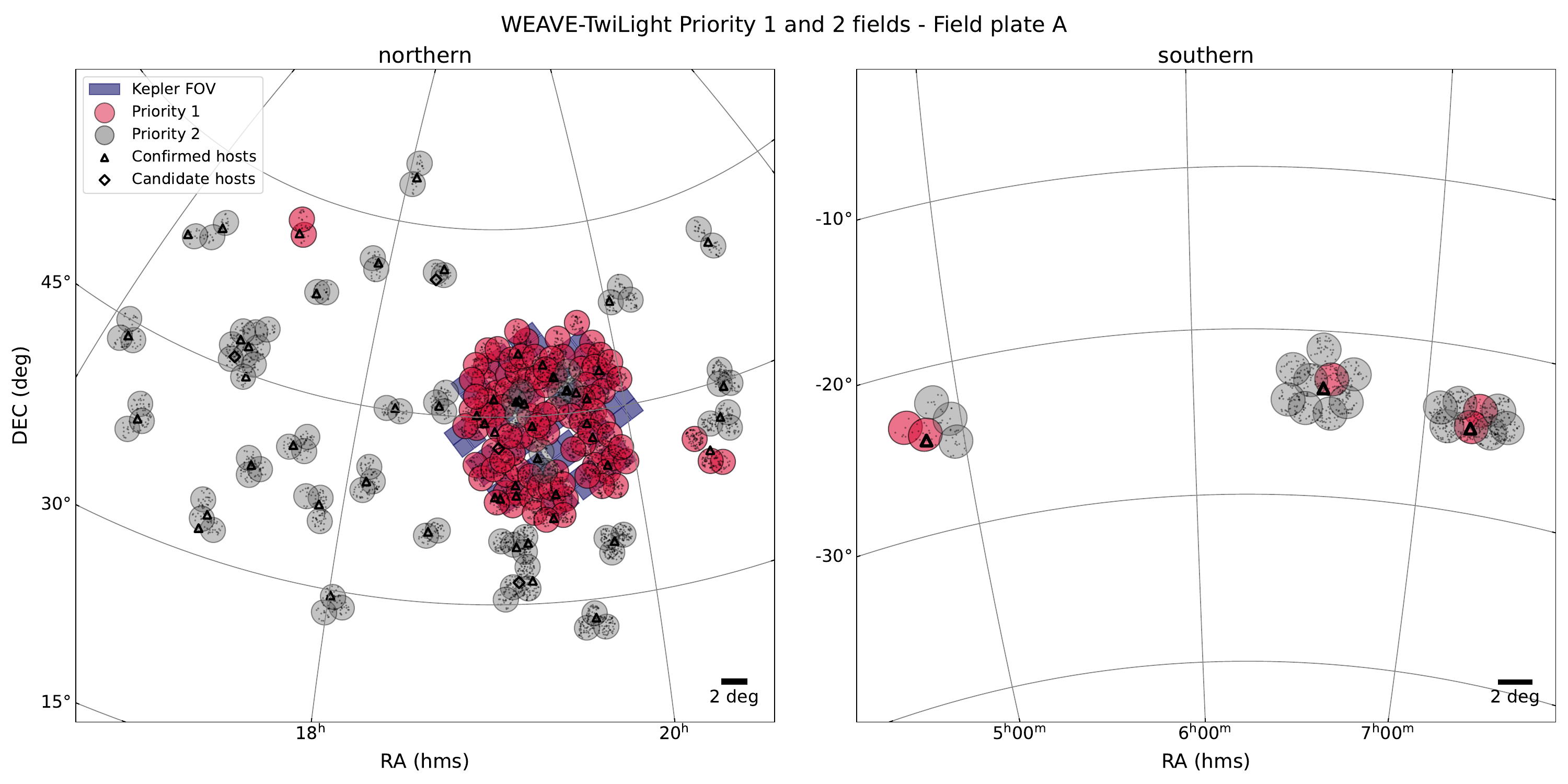}
   \caption{The northern (\textit{left} panel) and southern (\textit{right} panel) WEAVE-TwiLight fields derived for WEAVE's field plate A (on-sky projection in RA, Dec coordinates). 1st and 2nd priority classes, as defined in Section \ref{s:WTL_fields} are shown as \textit{red} and \textit{grey} circles, respectively. Known planetary host stars (NASA Exoplanet Archive and TESS Objects Of Interest) are depicted as \textit{black} triangles, candidate hosts as \textit{black} diamonds. To ease orientation, the Kepler field-of-view is shown in \textit{dark blue}.}
  \label{im:WTL_NS_A}
\end{figure*}
\begin{figure*}
\centering
 \includegraphics[width=17cm]{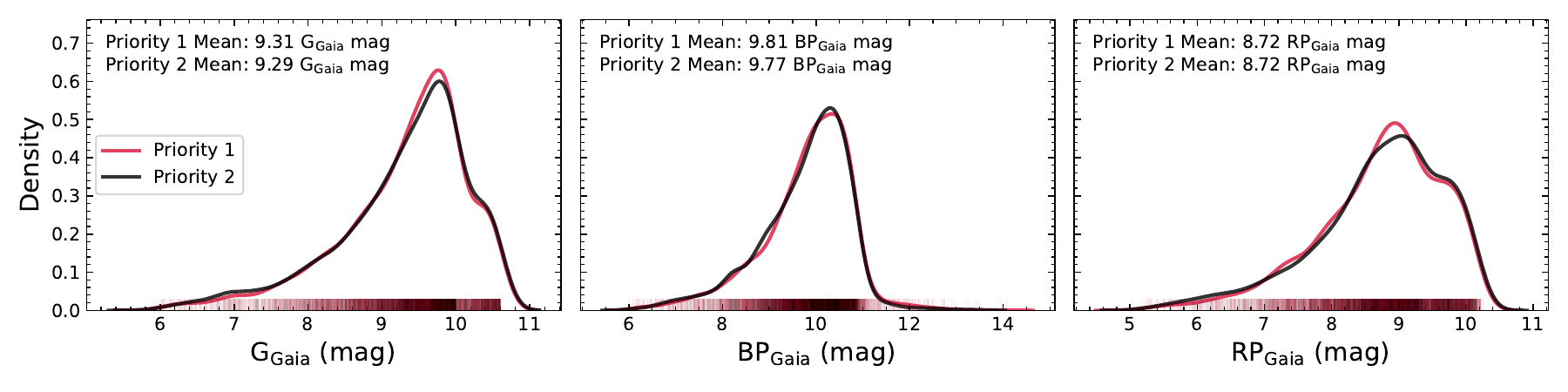}
   \caption{\textit{Gaia} magnitude distributions (KDE) for the WEAVE-TwiLight-Survey. Priority 1 and 2 targets (see Section \ref{s:WTL_fields}) are shown as \textit{red} and \textit{black} curves, respectively. The vertical lines on the bottom illustrate individual magnitude values of stars in the sample.}
  \label{im:mag_distrib}
\end{figure*}

\section{On-sky fields for the WEAVE-TwiLight-Survey} \label{s:WTL_fields}
\begin{table}
    \centering
    \begin{tabular}{lcccc} \toprule
         Priority Class & 1& 2& 1 + 2\\ \midrule
         Fields Plate A&104&101&205\\ \midrule
         \# north &  3\,127 & 2\,342 &5\,469\\
         \# south& 170& 645 &815\\ \midrule
         Total nr. of targets&&&6\,284 \\\midrule \midrule
         Fields Plate B&105&106&211 \\ \midrule
         \# north &  3\,177 & 2\,479 &5\,656\\
         \# south& 122& 514 &636\\ \midrule
         Total nr. of targets&&&6\,292 \\ \bottomrule \\
    \end{tabular}
    \caption{Number of fields and science targets in each priority class for the northern and southern WTLS fields.The varying number of allocated science targets between field plates reflect the guide fibre availability at the time the WTLS fields were defined. Since the operational guide fibres occupied different positions on each plate, some field batches provide sufficient guide fibre coverage for one plate but not necessarily for the other.}
    \label{t:num_science_targets}
\end{table}

As mentioned above, at the time the final WTLS fields were defined both WEAVE field plates had seven operational guide fibres. Since each field within a field-batch requires two off-centre guide fibres and one central guide fibre, a total of seven guide fibres allows three separate fields to be observed with the same fibre configuration. Eight guide fibres do not permit an additional field due to the constraints described above, while six guide fibres would limit each batch to only two fields. However, the overall number of fields with sufficient guide fibre allocation increases with the number of guide fibres in operation, as shown in Figure \ref{im:density_guide_allocation}. This is because a greater number of available guide fibres allows for more geometric configurations in which guide stars fall within fibre reach (see Fig.~\ref{im:guide_fibre_alloc}). In our case, the method produced $1\,400$ fields with sufficient guide star coverage (see Fig.~\ref{im:density_guide_allocation}, panel c), containing $\sim 60\,000$ potential science targets. \par

This number is much larger than what can realistically be observed during a WEAVE open-time program, prompting us to select only field-batches that contain at least one known exoplanet host and a selection of additional targets within the Kepler field (see Fig. \ref{im:WTL_NS_A}). These were then further split into the following two priority classes:
{
\renewcommand{\labelenumi}{\arabic{enumi}}
\begin{enumerate}
    \item - field-batches with targets observed by APOGEE
    \item - batches containing targets without APOGEE coverage
\end{enumerate}
}

WTLS will be among the first high-resolution MOS surveys conducted with WEAVE. Therefore, priority has been given to stars with APOGEE coverage to enable direct comparisons with existing data.
This will extend the wavelength coverage into the optical for a large fraction of the targets and support internal calibration efforts for both WEAVE and \textit{PLATO}. In addition to the $1\,279$ WTLS targets with APOGEE coverage, we cross matched our catalogue with stars observed by MELCHIORS ($11$ targets; \citealp{royerMELCHIORSMercatorLibrary2024}), LAMOST ($583$ targets; \citealp{zhaoLAMOSTSpectralSurvey2012}), and RAVE ($118$ targets; \mbox{\citealp{steinmetzRadialVelocityExperiment2006}}). These overlaps will further aid the validation efforts of the WEAVE stellar-analysis pipeline. It should be noted that in the southern portion of the fields accessible to the WHT, no stars with APOGEE coverage could be identified. It was therefore decided to assign priority 1 to all batches containing planetary hosts and priority 2 to the remainder. \par

The final science target yield per priority class after applying the above selection cuts is listed in Table \ref{t:num_science_targets}. It is evident that the number of targets varies slightly depending on the field plate to which the fields have been allocated. This is due to the fact that the positions of the inoperative guide fibres are different for each field. This means that some guide stars accessible to the guide fibres of plate A cannot be reached by the guide fibres of plate B. To allow maximum flexibility during observations, OBs with configured fibres for both field plates will be submitted to the telescope. For plate A, Figure \ref{im:WTL_NS_A} shows the final fields for WEAVE-TwiLight as red and grey circles for the priority classes 1 and 2, respectively. Known exoplanet hosts in the sample are indicated by black triangles, while candidate hosts are shown as diamonds. The field distribution on the sky for plate B is illustrated in Figure \ref{im:WTL_NS_B}. To aid orientation, the \textit{Kepler} field-of-view is shown in dark blue.
Distributions (Kernel Density Estimates, KDE) of \textit{Gaia} magnitudes for WTLS fields are plotted in Figure \ref{im:mag_distrib}. In total, the final fields contain $62$ confirmed exoplanet host stars. \par
This provides a solid foundation for future efforts to characterize thousands of exoplanet hosts, including targets in the TESS continuous-viewing zone and those expected from the upcoming 4th \textit{Gaia} data release \citep{sozzettiGaiaAstrometryExoplanetary2023}. Over time, this will yield a robust spectroscopic dataset, enabling detailed studies of chemical abundance patterns linked to planet formation. \newline

\begin{figure*}
\centering
 \includegraphics[width=17cm]{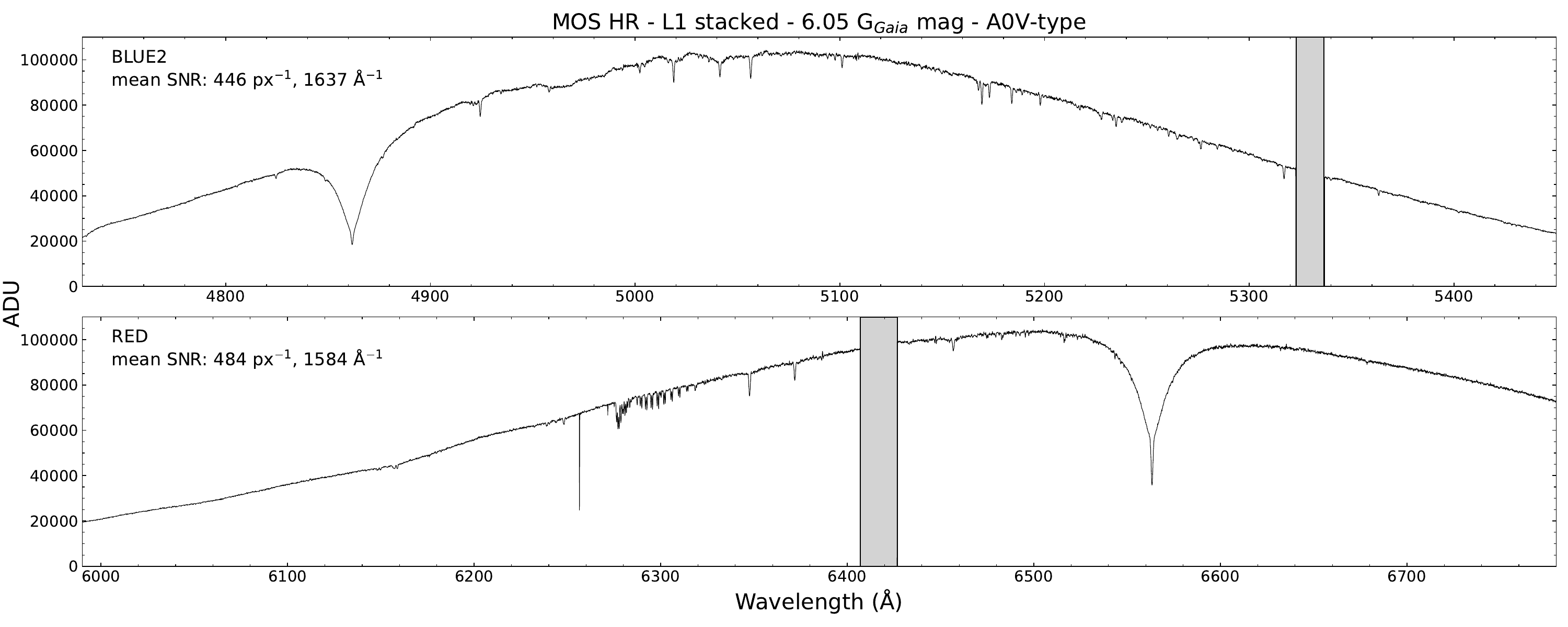}
   \caption{\textit{Blue2} (\textit{top}) and \textit{red} (\textit{bottom}) arm MOS HR data of a $6.05~G_{\mathit{Gaia}}$ mag, A0V-type star observed with WEAVE. The mean SNR (per pixel and per Angstrom) is given in the respective panels. Grey areas depict coverage gaps due to projection effects onto the CCDs. The total exposure time was $300~\si{\second}$.}
  \label{im:spectrum_6mag}
\end{figure*}
\begin{figure*}
\centering
 \includegraphics[width=17cm]{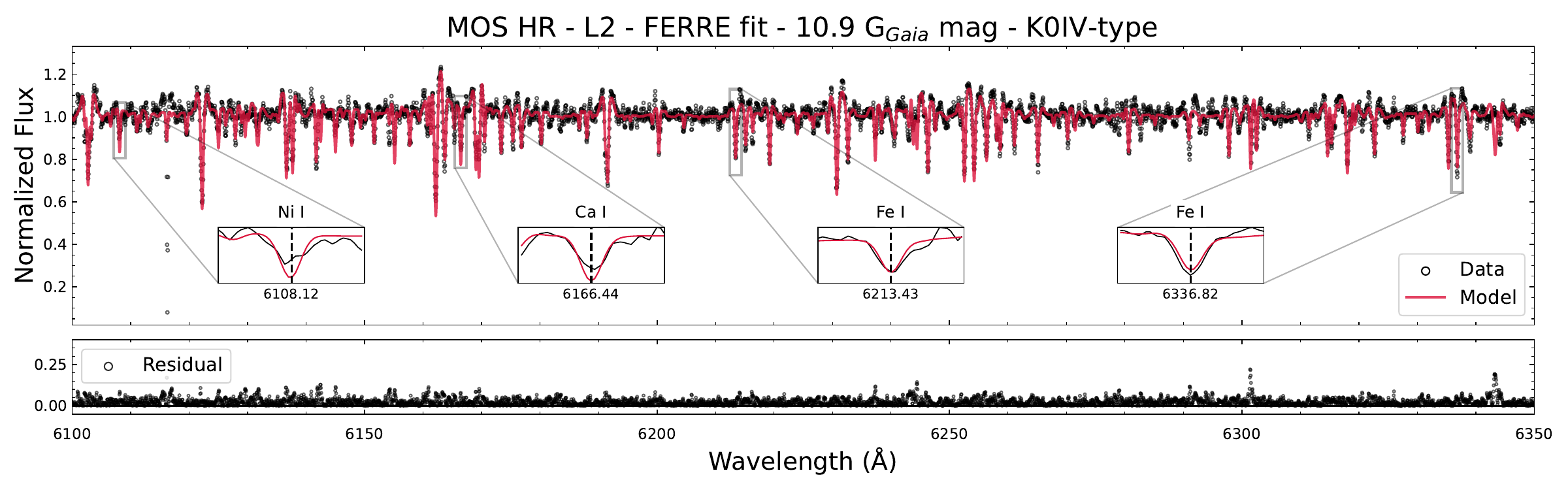}
   \caption{Red arm spectrum (zoomed in) of a K0IV-type star, overlaid with the best-fit atmospheric model (\textit{red}), determined using \textsc{ferre} \mbox{\citep{allendeprietoSpectroscopicStudyAncient2006}}. Residuals are shown in the \textit{lower} panel. Inlays depict spectral features of Ni I, Ca I and Fe I at their respective wavelengths.}
  \label{im:model_obs_spectrumFR}
\end{figure*}

To enable this, WTLS will use WEAVE's \textit{HR blue2} mode, providing good coverage of the key Magnesium I triplet lines (5172.68 Å, 5183.60 \AA). Mg is a near-pure $\alpha$-element produced predominantly in core-collapse supernovae. As a result, high [Mg/Fe] ratios are characteristic of older, $\alpha$-enhanced thick-disk stars, whereas lower [Mg/Fe] values are typical of younger thin-disk populations (e.g. \citealp{mashonkinaAbundancesAProcessElements2019}, \citealp{tsantakiArielStellarCharacterisation2025}). These population differences are reflected in exoplanet properties. For example, \cite{adibekyanStellarPlanetaryComposition2015} note that low-mass planet hosts show a tendency toward a higher [Mg/Si] ratio, compared to stars without planets, after controlling for overall metallicity trends. More recently, \cite{behmardLinkRockyPlanet2025} found an inverse correlation between host star [Mg/Fe] and rocky planet density, with low-[Mg/Fe] (Fe-rich) stars forming higher-density planets (larger iron cores) and high-[Mg/Fe] thick-disk stars yielding lower-density, less iron-rich planets. However it should be noted that their sample consists of 22 stars, and the inverse correlation seen in their study is yet to be confirmed in larger datasets. This trend, if confirmed, is anchored by thick-disk hosts, indicating that Mg abundance patterns reflect the refractory composition of the forming planetary core and are consistent with either core-accretion or pebble-accretion pathways in metal-rich environments. Indeed, \cite{tsantakiArielStellarCharacterisation2025} confirm that giant planets occur preferentially around thin-disk, metal-rich (low-[Mg/Fe]) stars, whereas $\alpha$-enhanced stars in the thick disk more often host only low-mass rocky planets. Other elements (e.g. Ca, Si, Fe) also contribute to planet interiors, but Mg provides a particularly clear signal because of its single nucleosynthetic origin, reducing degeneracies in interpretation. When assessing planet populations, the Galactic chemical context must not be ignored, as there are signs that planet formation is following the Galactic chemical evolution \citep{teixeiraWhereMilkyWay2025}. Systematic differences in elemental ratios (e.g. [Mg/Fe], [C/O]) between thin- and thick-disk populations have to be taken into account in order to avoid biases in planet characterization (e.g. \mbox{\citealp{dasilvaArielStellarCharacterisation2024}}).

To address the above and enable consistent analysis, we aim to derive stellar parameters ($T_{\mathrm{eff}}$, log$g$, $v_{\text{micro}}$, [Fe/H], [$\alpha$/H]) and twelve chemical abundances (V, Ni, Mg, Ca, Si, Na, Al, Ba, Sr, Cu, O and C) for all observed targets. Additional elements are likely to be included in future updates.
Work is currently underway to integrate a robust chemical abundance pipeline into WEAVE's Advanced Processing System (APS) as part of ongoing efforts to incorporate a number of contributed software packages from WEAVE Science Team members. \par
WTLS will be a unique, magnitude-limited multi-object spectroscopic survey designed to compare the chemical abundance patterns of confirmed exoplanet hosts with the broader F, G, K and A star population. The resulting dataset will be highly homogeneous and comparable in size to other studies exploring links between host stars and their planets (e.g. \citealp{clarkGALAHSurveyImproving2022}; \citealp{yunConnectionsPlanetaryPopulations2024a}).

\section{Initial test observations}\label{s:test_observations}
In preparation for WTLS, test observations were successfully carried out between 27 August and 10 September 2024 and from 26 to 27 February 2025. In this context, several OBs were observed during twilight hours, testing merged field configurations and pipeline handling for targets with magnitudes between \mbox{$6 \leq V \leq 11$}. For these tests, individual OBs with $\sim 20$ targets per field were created, to ease manual assessment in terms of correct fibre placement in between pointings. Note that these fields were not part of the final priority fields as defined in Section \ref{s:WTL_fields}, although a small overlap with a minor fraction of WTLS stars does exist. In total, spectra from 98 stars have been recorded. Figure \ref{im:spectrum_6mag} shows a typical data product after wavelength calibration, sky subtraction, and stacking, for an A0V-type star with $G_{\mathit{Gaia}} = 6.05$ mag. The plot shows a MOS HR observation with the \textit{blue2} (top) and \textit{red} (bottom) spectrograph arms. The 2024 test observations were limited to a total exposure time of $120\si{\second}$, whereas the 2025 runs used a longer integration time of $t_{\mathrm{exp}} = 300\si{\second}$, resulting in an increase in signal-to-noise ratio of $\sim1.6$. Preliminary processing runs, testing both CPS and APS pipelines, have been completed successfully, demonstrating that twilight OBs can be handled by the system. A zoomed-in comparison between a K0IV-type stellar spectrum taken with the red arm and a stellar atmospheric model is shown in Figure \ref{im:model_obs_spectrumFR}. The fit was enabled using the $\chi^2$ minimization code \textsc{ferre}\footnote{\url{https://github.com/callendeprieto/ferre}} \mbox{\citep{allendeprietoSpectroscopicStudyAncient2006}} as part of the APS stellar parameter pipeline. Figures~\ref{im:spectrum_6mag} and \ref{im:model_obs_spectrumFR} both show that the new mode is feasible and results in high-quality data. 
A comprehensive description of the survey design and an initial analysis of a sample subset will be presented in a forthcoming paper (Hajnik et al., in prep).

\section{Conclusion}\label{s:conclusion}
We have presented a novel approach that allows observations of low-target-density fields with WEAVE, while significantly reducing the observational overhead that could be introduced by such fields. The WEAVE-TwiLight-Survey (WTLS) will act as the pilot for this new mode, which is enabled by superimposing multiple observational fields (pointings) onto the same fibre configuration. These fields can then be observed in quick succession without tumbling the field plates or requiring new calibration exposures in between. \par

The starting points for WTLS fields are the \textit{PLATO} long-duration-phase fields (LOPN/LOPS, \citealp{nascimbeniPLATOFieldSelection2022, nascimbeniPLATOFieldSelection2025}), which are further enhanced by adding bright stars from \textit{Gaia} DR3. We introduced a field definition method employing a heuristic approach to the geometric set cover problem, enabling an overall field coverage of $\sim97$ percent of our \textit{proto} catalogues. \par

To enable the successful assignment of guide fibres in the run-up to science fibre allocation, a custom algorithm has been developed. We find that its ability to allocate the necessary guide fibres for merged field configurations depends on the surface density of guide stars and the availability of operational guide fibres. On-sky plots of the final fields are presented in Figures~\ref{im:WTL_NS_A} and \ref{im:WTL_NS_B}, along with the science target yields for different priority classes and field plates (Table~\ref{t:num_science_targets}). \par

To test the feasibility of the merged field configurations, test observations were successfully performed in August (27, 28) and September (10) 2024 and February (26, 27) 2025. From these tests, we conclude that the sky subtraction of WEAVE spectra taken at twilight works well and that the data products of observations taken with the new mode are fit for scientific use. Additional tests focusing on large ($> 2^\circ$) offsets between merged fields have been carried out. At the time of writing, the largest successfully tested offset between fields is $\sim15^\circ$. \par

Changes to the telescope systems required to enable the new mode (WEAVE-TL) have now been implemented. The WEAVE-TwiLight-Survey was awarded 37.8 hours of observing time by the Panel for the Allocation of UK Telescope Time (PATT) for the 2025A2-2025B1 open-time cycle. An initial set of WTLS science OBs was observed in November 2025.

\section*{Acknowledgements}
We thank the anonymous reviewers for their detailed and constructive feedback, which has significantly improved the quality of the manuscript. We also thank the WEAVE consortium internal reviewer, Dr Nicolas Martin, for comments that further strengthened the paper. \newline

Based on observations made with the William Herschel Telescope operated on the island of La Palma by the Isaac Newton Group of Telescopes in the Spanish Observatorio del Roque de los Muchachos of the Instituto de Astrofísica de Canarias (WEAVE proposal WS2025A2-007). \newline

Funding for the WEAVE facility has been provided by UKRI STFC, the University of Oxford, NOVA, NWO, Instituto de Astrofísica de Canarias (IAC), the Isaac Newton Group partners (STFC, NWO, and Spain, led by the IAC), INAF, CNRS-INSU, the Observatoire de Paris, Région Île-de-France, CONACYT through INAOE, the Ministry of Education, Science and Sports of the Republic of Lithuania, Konkoly Observatory (CSFK), Max-Planck-Institut für Astronomie (MPIA Heidelberg), Lund University, the Leibniz Institute for Astrophysics Potsdam (AIP), the Swedish Research Council, the European Commission, and the University of Pennsylvania. The WEAVE Survey Consortium consists of the ING, its three partners, represented by UKRI STFC, NWO, and the IAC, NOVA, INAF, CNRS-INSU, INAOE, Vilnius University, FTMC – Center for Physical Sciences and Technology (Vilnius), and individual WEAVE Participants. Please see the relevant footnotes for the WEAVE website\footnote{\url{https://weave-project.atlassian.net/wiki/display/WEAVE}} and for the full list of granting agencies and grants supporting WEAVE\footnote{\url{https://weave-project.atlassian.net/wiki/display/WEAVE/WEAVE+Acknowledgements}}. JALA acknowledges financial support provided by the Spanish Ministerio de Ciencia, Innovación y Universidades (MICIU) through the project PID2023-153342NB-I00. \newline

This work has made use of data from the European Space Agency (ESA) mission
{\it Gaia} (\url{https://www.cosmos.esa.int/gaia}), processed by the {\it Gaia}
Data Processing and Analysis Consortium (DPAC,
\url{https://www.cosmos.esa.int/web/gaia/dpac/consortium}). Funding for the DPAC
has been provided by national institutions, in particular the institutions
participating in the {\it Gaia} Multilateral Agreement. \newline

This project has received funding from UKRI Horizon Europe Guarantee (grant no. EP/X033066/1).
This project has received funding from the European Union's Horizon 2020 research and innovation
programme under the Marie Skłodowska-Curie grant agreement No 101072454 (MWGaiaDN)\footnote{\url{https://cordis.europa.eu/project/id/101072454}}.
PB acknowledges support from the ERC advanced grant N. 835087 -- SPIAKID.  D.A. acknowledges financial support from the Spanish Ministry of Science and Innovation (MICINN) under the 2021 Ram\'on y Cajal program MICINN RYC2021‐032609. This project has received funding from the European Research Council (ERC) under the European Union’s Horizon 2020 research and innovation programme (Grant agreement No. 101020057).

\section*{Conflict of Interest}
The authors declare no conflict of interest.

\section*{Data Availability}
The fully reduced data for the WEAVE-TwiLight-Survey will be available from the WEAVE Archive System (WAS). Custom algorithms enabling the new mode will be publicly released once integration into the WEAVE workflow is complete. Until then, the authors will provide the field-definition algorithms for WEAVE-TL upon reasonable request.



\bibliographystyle{rasti.bst}

\bibliography{main.bbl} 



\onecolumn
\appendix
\section{}
\subsection{WEAVE-TwiLight fields for plate B}
\vspace{1em}  

\centering
\includegraphics[width=17cm]{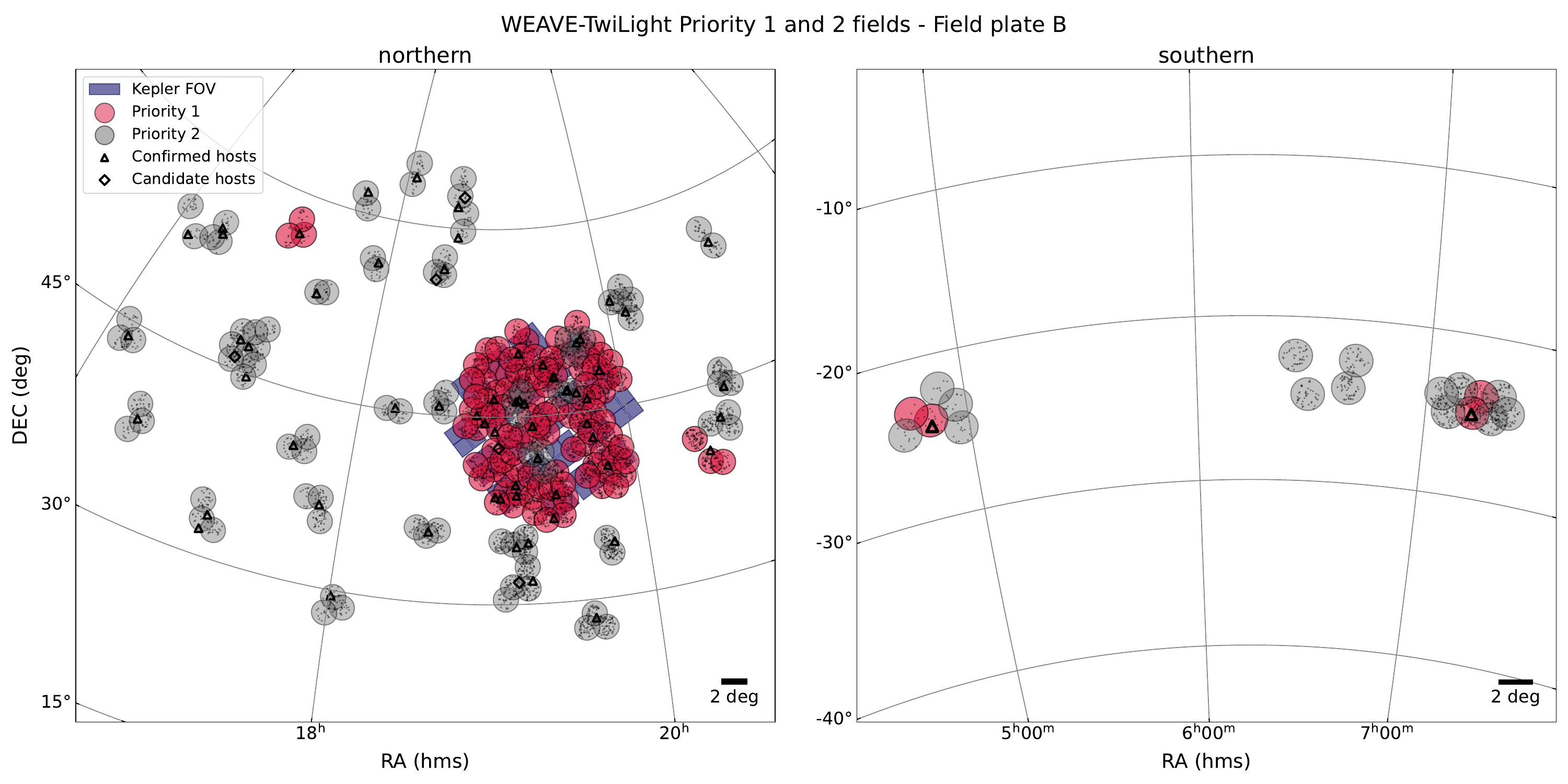}

\vspace{0.5em}  
\noindent
\textbf{Figure \ref{im:WTL_NS_B}.} Same as Figure \ref{im:WTL_NS_A} but for fields allocated to plate B.
\label{im:WTL_NS_B}


\label{lastpage}
\end{document}